TITLE PAGE

**Improving Hypertension and Diabetes Outcomes with Digital Care Coordination and Remote Monitoring in Rural Health**

Corresponding Author:
Katherine K. Kim, PhD, MPH, MBA
University of California Davis
School of Medicine, Department of Public Health Sciences
One Shields Avenue
Davis, California 95616 USA
kathykim@ucdavis.edu

Scott P. McGrath, PhD
University of California Berkeley
Center for Information Technology Research in the Interest of Society (CITRIS)
330 Sutardja Dai Hall
Berkeley, CA 94720 USA
smcgrath@berkeley.edu

David Lindeman, PhD, MSW
University of California Berkeley
Center for Information Technology Research in the Interest of Society (CITRIS)
330 Sutardja Dai Hall
Berkeley, CA 94720 USA
dlindeman@berkeley.edu

**ABSTRACT**

Chronic illnesses are a global concern with essential hypertension and diabetes mellitus among the most common conditions. Remote patient monitoring has shown promising results on clinical and health outcomes. However, access to care and digital health solutions is limited among rural, lower-income, and older adult populations. This paper repots on a pre-post study of a comprehensive care coordination program including connected, wearable blood pressure and glucometer devices, tablets, and medical assistant-provided health coaching in a community health center in rural California. The participants (n=221) had a mean age of 54.6 years, were majority female, two-thirds spoke Spanish, 19.9% had hypertension, 49.8% diabetic, and 30.3% both conditions. Participants with hypertension achieved a mean reduction in systolic blood pressure of 20.24 (95% CI: 13.61, 26.87) at six months while those with diabetes achieved a mean reduction of 3.85 points (95% CI: 3.73, 4.88). These outcomes compare favorably to the small but growing body of evidence supporting digital care coordination and remote monitoring. These results also support the feasibility of well-designed digital health solutions yielding improved health outcomes among underserved communities.

# INTRODUCTION

Chronic conditions are a global concern with diabetes mellitus[1] and essential hypertension[2] among the highest prevalence conditions.[3,4] For adults served by the United States (US) Medicaid program which covers healthcare for people with low income and limited resources, diabetes and hypertension are among the ten most prevalent and highest cost chronic conditions.[5] Minority populations in the US including Black, Latine, Asian, as well as rural residents experience disproportionately higher rates of hypertension[6,7] and diabetes[8-10] than White and urban. Latine populations in particular have persistent challenges with uncontrolled hypertension, worse glycemic control, and higher death rates.[11]

The geographic areas in the US with the highest prevalence of chronic conditions also experience the challenges associated with social drivers of health that include socioeconomic disadvantages and barriers to healthcare access.[12] These communities often face multifaceted barriers including geographic isolation, healthcare provider shortages, and inadequate healthcare infrastructure, which contribute to healthcare disparities.[13] Residents in rural areas experience negative impacts such as delayed diagnoses [14], treatment gaps [15], and overall poorer health outcomes in comparison to their urban counterparts.[16]

California's rural and agricultrual Central Valley, has a predominantly Latine population facing the state's worst environmental conditions, extreme poverty, and health professional shortages. Before the pandemic, community health centers nationwide, including California, were not authorized for telehealth visits. They struggled to meet virtual visit demands and monitor chronic illnesses during the pandemic due to disparities and limited technology access.[17] There remains a gap in understanding how to address the challenges and barriers to implementing and adopting telehealth and RPM in these communities.

There is some evidence that telehealth technologies including remote patient monitoring (RPM) may be a promising strategy for disease management in a variety of populations.[18] RPM has shown promising results on clinical and health outcomes in a small number of papers and in management of hypertension and diabetes. A meta-analysis of 27 studies using biosensors for hypertension reported an average systolic blood pressure (SBP) reduction of 2.62 mmHg (95% CI: −5.31, 0.06) and diastolic blood pressure (DBP) reduction of 0.99 mmHg (95% CI: −2.73, 0.74).[19] Another meta-analysis found more pronounced reductions, with an average SBP reduction of 7.07 mmHg and DBP reduction of 3.11 mmHg across 18 studies.[20] In a meta-analysis of 20 studies of diabetes remote monitoring found an average reduction in HbA1c of 0.55% compared to usual care (Range: -1.2%, -0.1%).[21] Another review showed that RPM systems notably reduce glycosylated hemoglobin levels in type 2 diabetes patients, with a mean reduction of −0.32 (95% CI:−0.45, −0.19).[22]

Early research indicates promise in populations experiencing health disparities [23,24] but progress is hampered by issues such as limited internet connectivity, digital literacy barriers [25], and cultural variations in healthcare perception and utilization. [26] In California's rural and agricultural Central Valley, the population is predominantly Latine. In addition to documented health disparities, there are high rates of poverty and limited technology access.[17,27] The ACTIVATE program was conducted in this particular setting to understand how to develop and implement a digital health solution to address these challenges.

# METHODS

## Study Design and Setting

This study is a pre-post study of RPM delivered in a primary care team-based model with health coaching in one federally qualified health center (FQHC) which serves all individuals without regard to ability to pay. The service area covers two counties in a primarily rural and agricultural region called the Central Valley of California.

## Recruitment

Eligibility criteria including adult patients (18 years of age or older) of the health center with a diagnosis of diabetes mellitus or essential hypertension who had at least one encounter within the previous two years. For those with diabetes mellitus the additional inclusion criteria were the most recent hemoglobin A1c value was equal to or greater than 8.0. For hypertension inclusion, the systolic blood pressure was equal to or greater than 140 mg or the diastolic blood pressure was equal to or greater than 80 mg/hg. Patients with end-stage or advanced disease as determined by the provider at the health center were excluded. Participation was limited to patients speaking English or Spanish because the health care teamlet, composed of a provider, health coach, and digital health navigator, was bilingual in those languages. A list of eligible patients was created and contacted sequentially by the health center provider, licensed vocational nurse/health coach (LVN), or community health worker/health coach (CHW) until a target number of program participants was reached. An enrollment survey was administered orally by the LVN or CHW. The survey included demographics, technology access, digital literacy, health literacy, and diabetes and hypertension self-efficacy instruments. Recruitment occurred from April 2021 to December 2022, with follow-up data collection until December 2023.

**Digital Care Coordination Program**

A technology kit was provided to patients including connected blood glucose and blood pressure monitors, a data-enabled tablet, virtual health coaching sessions twice per month, and virtual or in-person provider visits as needed. Participants were provided with RPM devices, OneTouch Verio Flex glucometer and/or Omron 7 or 10 Series blood pressure monitor, and a Samsung Galaxy Tab S6 LITE tablet with a data plan if needed. Participants used a custom app to pair RPM devices via Bluetooth and transmit data to the ACTIVATE system. A clinic community outreach worker filled the role of digital navigator and assisted participants with technology setup and use either in the clinic or by phone. Health coaches who were trained medical assistants reviewed RPM data through a secure web portal. Weekly summaries of each participant's RPM data were uploaded to the electronic health record (EHR). Participants were discussed in the weekly team "huddle," a 15-30 meeting with provider, medical assistant/health coach and digital navigator. The discussions included review of RPM data trends, patient goals and challenges, and any appropriate care plan adjustments. In depth description of the community co-design methods applied to development of the platform, and details of the digital care coordination program are described in a previous paper.[28]

**Study Data**

Participant demographics and pre-post A1c and blood pressure measures were abstracted by the clinic staff from the EHR. RPM data collected electronically from devices and transmitted in real-time to ACTIVATE included timestamps, glycated hemoglobin (eAG) and blood pressure and heart rate. The data collected in the enrollment survey included validated instruments (Table 1).

**Table 1. Data collection instruments and score interpretations**

| Instrument | Domains | Score | Interpretation |
|---|---|---|---|
| PROMIS Self-Efficacy for Managing Chronic Conditions[29,30] | 11 items across 5 domains with total and domain subscores.[a]<br>• Managing daily activities<br>• Managing emotions<br>• Medications and treatments<br>• Social interactions<br>• Symptoms | $\geq 60$<br>$41 - 59$<br>$\leq 40$ | High self-efficacy<br>Average<br>Low |
| Summary of Diabetes Self-Care Activities (SDSCA)[31,32] | 11 questions regarding self-care with subscores for the following.<br>• General Diet: Average number of days per week following a healthy eating plan | $\geq 5$ days/wk<br>3-4 days/wk<br>$\leq 2$ days/wk | High adherence<br>Moderate<br>Low |
| | - Specific Diet: Number of days consuming at least five servings of fruits/vegetables and avoiding high-fat foods (reverse scored)
- Exercise: Number of days engaging in at least 30 minutes of physical activity
- Blood Glucose Testing: Number of days performing recommended glucose testing
- Foot Care: Number of days inspecting feet and shoes
- Smoking Status: Binary (0 = non-smoker, 1 = smoker) | | |
| Hypertension Self-Care Activity Level Effects (H-Scale)[33] | 31 items with measures for hypertension self-care activities with each subscale scored. | | |
| | - Medication Adherence (Score 0-21) | 21
$\leq 14$ | Full adherence
Inconsistent |
| | - Diet - Dietary Approaches to Stop Hypertension Questionnaire (DASH-Q)[34] (Score 0-77) | $\geq 59$
39-58
$\leq 38$ | High diet quality
Medium
Low |
| | - Physical Activity (Score 0-14) | $\geq 8$ | Recommended adherence |
| | - Smoking Cessation | 0
1 | Smoker
Non-smoker |
| | - Weight Management: engaging in weight-conscious behaviors. | $\geq 5$ days/wk | Adherence |
| | - Alcohol Consumption | $\leq 2$ drinks/day for men
$\leq 1$ drink/day for women | Adherence |

[a] Each domain is scored separately, and reported as T-scores to facilitate interpretation. PROMIS instruments utilize item-level calibrations, which means that each question (item) is weighted based on its statistical properties derived from Item Response Theory (IRT). This approach ensures that scores are not simply summed raw values but instead reflect a standardized measure of self-efficacy across diverse populations. The responses to individual items are converted into a T-score metric where the mean = 50 and SD = 10. Scores were calculated using HealthMeasures.net, the official scoring platform for PROMIS instruments, which applies the IRT-based algorithms to ensure standardized results.

**Analysis**

Descriptive statistics outlined the demographic and baseline characteristics of the study population at the single site. Longitudinal changes in A1c and blood pressure were analyzed at three- and six-months post-enrollment, comparing these measures to pre-enrollment baselines. A1c Calculation: The pre-enrollment A1c value was obtained from the electronic health record (EHR) for each participant, and average glucose readings transmitted via the ACTIVATE platform were converted to A1c using the ADA's formula eAG was converted to A1c using the American Diabetes Associations formula $eAG + 46.7/28.7 = A1c$ [35], with changes from baseline determined for the three and six-month periods. Blood Pressure Calculation: The pre-enrollment blood pressure value was obtained from the EHR for each participant, with average blood pressure readings calculated over months 3 and 6 and changes from baseline determined for these periods. Descriptive statistics were used to describe technology use.

**RESULTS**

Among 221 participants who initially enrolled, 195 attended the initiation meeting at which they received their RPM device and/or tablet and completed setup (See Figure 1). Of these initiators, 154 (70.9%) uploaded any data, 106 used their devices for at least three months, and 86 continued for six months. Over half the participants were female and the mean age was 54.6 years (See Table 2). The majority of participants (68.6%) spoke Spanish. About half of participants had a diagnosis of diabetes, one-fifth hypertension, and the remainder had both conditions. While almost all had a cell phone, only 64% had internet at home, and almost half of those respondents reported problems with internet always, often, or sometimes.

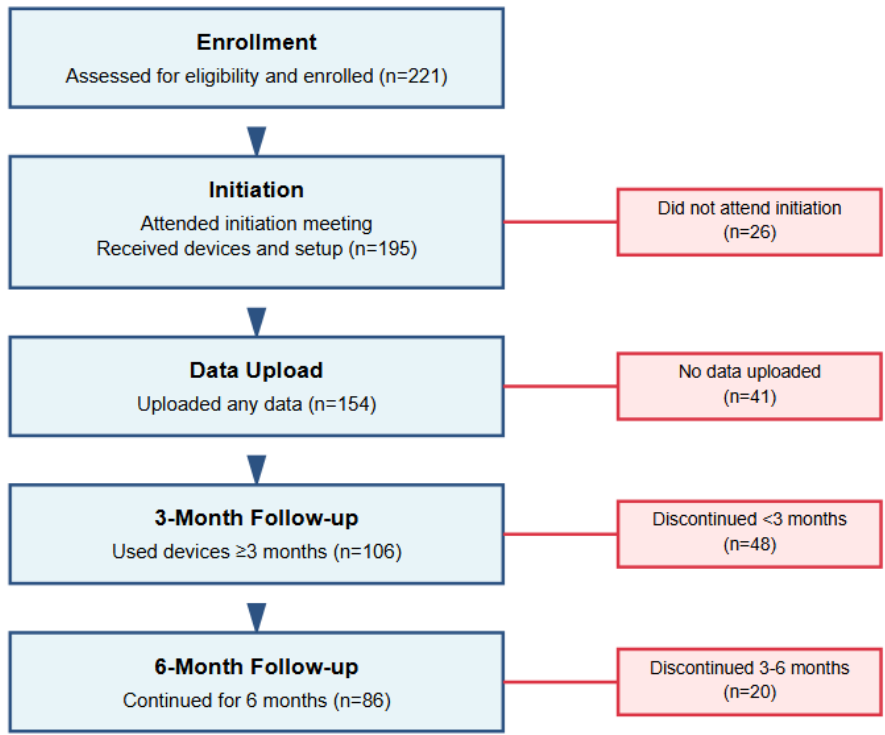

Fig. 1 Participant Retention

**Table 2. Participant Characteristics (n=221)**

| Characteristic | Number | Percent |
|---|---|---|
| Sex at Birth (n=221) | | |
|     Female | 138 | 62.4% |
|     Male | 83 | 37.6% |
| Age in Years (n = 221) | | |
| Mean (Range) | 54.5 (29-84) | |
| Race and Ethnicity (n=221) | | |
|     Hispanic or Latino | 194 | 87.8% |
|     White (not Hispanic or Latino) | 18 | 8.1% |

| | | | |
|---|---|---|---|
| Black or African American | | 5 | 2.3% |
| Asian | | 2 | 0.9% |
| Other Pacific Islander (Not Hawaiian) | | 1 | 0.5% |
| Declined to say | | 1 | 0.5% |
| Primary Language (n = 221) | | | |
| Spanish | | 150 | 67.9% |
| English | | 70 | 31.7% |
| Portuguese | | 1 | 0.5% |
| Diagnosis (n = 221) | | | |
| Diabetes Mellitus (all types) | | 110 | 49.8% |
| Essential Hypertension | | 44 | 19.9% |
| Both | | 67 | 30.3% |
| Technology Experience | | | |
| Do you have a cell phone? (n=208) | Yes | 195 | 93.8% |
| Do you have a tablet or iPad? (n=208) | Yes | 20 | 9.6% |
| Do you have a computer or laptop? (n=208) | Yes | 30 | 14.4% |
| Do you have an email account? (n=208) | Yes | 119 | 57.2% |
| Do you have access to the Internet where you live? (n=208) | Yes | 133 | 63.9% |
| If yes, how often do you have problems with your internet? (n=104) | | | |
| | Always | 3 | 2.9% |
| | Often | 5 | 4.8% |
| | Sometimes | 41 | 39.4% |
| | Never | 55 | 52.9% |

Participants overall rated their self-efficacy in managing chronic conditions as average for all domains and most were non-smokers. (See Table 3) Those with hypertension reported full adherence to alcohol recommendations but most did not adhere to weight management, medication management or physical activity recommendations. In addition, they had low diet quality. Those with diabetes reported moderate adherence on all SDSCA domains.

**Table 3. Participant Health Self-Efficacy on Enrollment**

| Assessment Instrument/Subscale | Score mean (SD) |
|---|---|
| PROMIS Self-Efficacy for Managing Chronic Conditions (n=208) | |
|     Managing Daily Activities | 48.4 (7.2) |
|     Managing Emotions | 48.7 (9.1) |
|     Managing Medications and Treatment | 46.4 (8.9) |
|     Managing Social Interactions | 48.4 (8.2) |
|     Managing Symptoms | 50.5 (3.53) |
|     Averaged Total | 49.11 (6.26) |
| Hypertension Self-Care Activity Level Effects (H-SCALE) (n = 180) | |
|     Medication adherence (n=125), Full adherence | 35% (44) |
|     DASH-Q (Diet Quality, n=126) | |
|         Low | 80.09% (100) |
|         Medium | 19.46% (25) |
|         High | 0.01% (1) |
| Physical Activity (n=121), Full adherence | 26.7% (48) |
| Smoking (n=180), Non-smokers | 93.3% (168) |
| Weight Management (n=185), Full adherence | 0% (0) |
| Alcohol consumption (n=178), Full adherence | 98.3% (175) |
| Summary of Diabetes Self-Care Activities (SDSCA) (n=183) | |
| General Diet | 3.5 (2.4) |
| Specific Diet | 4.0 (1.6) |
| Exercise Score | 3.0 (2.7) |
| Blood-Glucose Testing | 3.4 (3.0) |
| Foot Care | 3.7 (3.3) |
| Non Smoking Status (n=183)    % (n) | 171 (93.4%) |

The total number of measures transmitted over a six-month period was 35,264 with a mean of 229 uploads per person (SD = 284.42, range = 1 to 1,173). Hypertension participants had a mean of 152.98 uploads per person (SD = 175.16), and diabetes had a mean of 207.9 uploads per person (SD = 248.92).

Participants with diabetes started with a mean HbA1c of 11.22. (Table 4). At 3 months, they achieved a 3.44 point reduction (95% CI: 3.37, 4.35). At 6 months, they achieved a cumulative reduction of 3.85 points (95% CI: 3.73, 4.88), representing an additional 0.41 point reduction from 3 months. Among participants with hypertension, the average SBP was 157.94 and DBP was 83.36. These participants achieved a 16.49 reduction in SBP at 3 months (95% CI: 10.25, 22.73) and a total of 20.24 reduction in SBP at 6 months (95% CI: 13.61, 26.87), resulting in an average SBP of 138.36. DBP increased by 0.8 points at 3 months (95% CI: −3.93, 3.77) but decreased by 2.33 points at 6 months (95% CI: -2.30, 6.96), reaching an average of 81.05.

**Table 4. Health Outcomes**

| Diabetes | Hemoglobin A1c% mean (SD) | Hypertension | Blood pressure systolic, mm Hg mean (SD) | Blood pressure diastolic, mm Hg mean (SD) |
|---|---|---|---|---|
| Enrollment (n = 124) | 11.22 (1.87) | Enrollment (n = 41) | 157.94 (11.29) | 83.36 (6.67) |
| 3 mo (n = 124)[a] | 7.78 (1.81) | 3 mo (n = 41)[c] | 141.45 (17.29) | 83.43 (9.77) |
| 3-mo change (n = 124)[e] | 3.44 (2.75) | 3-mo change (n = 41)[e] | 16.49 (20.38) | -0.08 (12.58) |
| 6 mo (n = 76)[b] | 7.35 (1.49) | 6 mo (n = 31)[d] | 138.36 (16.18) | 81.05 (10.30) |
| 6-mo change (n = 76)[e] | 3.85 (2.54) | 6-mo change (n = 31)[e] | 20.24 (18.83) | 2.33 (13.15) |

Abbreviation: SD, standard deviation.
[a] Glucose readings over months 1 to 3 were averaged and converted to A1c using the ADA eAG to A1c conversion calculator.
[b] Glucose readings over months 4 to 6 were averaged and converted to A1c using the ADA eAG to A1c conversion calculator.
[c] Blood pressure measures were averaged over month 3.
[d] Blood pressure measures were averaged over month 6.
[e] Indicates reduction in measure.

**DISCUSSION**

This single site study (n=221) in a rural and agricultural community health center population suggests that digital care coordination and remote patient monitoring can yield improved health outcomes. The results are in line with the previous feasibility and pilot study in the same setting in which use of ACTIVATE (n=50) showed positive outcomes with reductions in SBP averaging 13.55 mm Hg (SD: 23.31) and a small reduction in (DBP) at 6 months as well as a decrease in A1c of 4.19 percentage points (SD: 2.69) at 6 months.[28]

These results add to the growing evidence base of previous studies of home-based RPM and telemonitoring of hypertension and diabetes with a variety of intervention designs. Related to hypertension RPM, Martin-Ibanez, et.al., analyzed the 24-month effects of home RPM and self-titration of medications, finding significant improvements of 3.4 mm Hg SBP and 2.5 mm Hg DBP.[36] Kim, et.al., conducted a three-arm RCT of RPM, with physician office follow-up visits, remote physician follow-ups, and usual care.[37] The authors found all reduced SBP and there was no significant difference between groups. In contrast, a three-arm RCT by Mehta, et.al., studied RPM with text-messaging for data collection, medication adherence, and feedback by Mehta, et.al., found no improvement in outcomes of RPM with and without family social support compared to usual care.[38] The interventions in these studies varied from use of support, text messaging, and physician office visits. A recent systematic review of 18 RCTs of hypertension RPM found an overall mean weighted decrease of 7.07 points (SBP) and 5.07 points (DBP) for the intervention group, significantly better than for control groups (mean change effect sizes of 1.1 SBP and 0.98 DBP).

Related to diabetes RPM interventions, Shane-McWhorter et al., report on a program similar to ACTIVATE in which diabetes and hypertension RPM was conducted in community health centers in a pre-post design (n=109).[39] The authors report a mean decrease in A1c of 1.92 and SBP decrease of 7.8. In an other study by Odom, et.al., employees with type 1 or type 2 diabetes (n=50) who underwent an intervention combining education, telehealth, medication management, and RPM, A1c was reduced by 1.8% after 2-year follow-up and there was a reduction in the cost of care.[40] Tang and colleagues' RCT of nurse-led diabetes RPM and educational intervention (n=showed 6-month improvement of intervention group (n=193) over the control (n=189), but the significant difference was not sustained at 12 months.[41] In a large, cluster-randomized pragmatic trial of several chronic conditions conducted in England, 513 of the 3,230 participants had type 2 diabetes.[42] The intervention which involved use of RPM and nurse care coordinators was associated with a small improvement in HbA1c from usual care (0.21% 95% CI, 0.04% to 0.38%).

Our study differs in several important ways from the other examples referenced in the discussion. First, ACTIVATE trained medical assistants employed at the clinic in health coaching to support patients using RPM. The medical assistants interacted with patients by phone or telehealth to offer health coaching based on patients' own health priorities, coordinate care as recommended by the provider, or to discuss the RPM data. Second, data were automatically transmitted in real-time from RPM device to the platform and the healthcare team held regular huddles to review the data and patient care plans. The results in our study for both patients with diabetes and hypertension are substantially better than most the literature, but we are not able to separate which aspects of the intervention design or component activities might have impacted outcomes. The effectiveness and cost-effectiveness of RPM may be impacted by the clinical context, capital investment, and organizational processes, underscoring the need for tailored implementation strategies.[22,43] Thus, a fruitful next step would include applying implementation science to understand potential best practices in multidimensional, complex interventions especially in in community settings. Some of the components to study might include: effectiveness of RPM with or without education or health coaching; actively collected RPM data collection compared to passively collected data from other wearables or sensors. the use of lay personnel such as medical assistants and community health workers for patient engagement and care coordination compared to licensed clinicians such as nurses and pharmacists.

A key challenge in ACTIVATE is the retention rate: 79% of initiators transmitted data, 56% completed 3 months monitoring, and only 39% completed 6 months monitoring. While the results for those who monitor are impressive there is much room for improvement if we those who discontinued prematurely had been retained. In contrast, the Shane-McWhorter and colleagues study retained 87.2% of participants in the 7-month program although their average improvement was much lower.[39] High attrition is a persistent challenge in the field.[18,19]  The field would benefit from better understanding of how best to recruit the individuals most likely to benefit from RPM and care coordination, and ways to improve participant engagement, particularly in light of barriers to digital health adoption for individuals of underserved communities.

### Limitations

This was a quality improvement study whose purpose was to offer an alternative to in-person visits for engaging patients in chronic disease self-management.  The protocol for the intervention was pragmatic and flexible in order to be responsive to participant preferences and needs. The participants were recruited from a convenience sample at one health center.  These factors limit the ability to understand causality or attribute the outcomes to specific intervention components. Caution should be used in generalizing study findings to the broader population.

## CONCLUSIONS

The ACTIVATE study suggests that digital care coordination is feasible and a potentially promising approach to chronic disease care in underserved communities. The intervention including RPM via connect blood pressure cuffs and glucometers delivered with health coaching and care coordination resulted in impressive improvement in A1c and blood pressure for participants who engaged in monitoring for at least three months. In addition to ACTIVATE's outcomes, the substantial usage of the technology by patients points to the interest in such interventions. In light of the challenges to digital health adoption in rural and low-resource communities, these results are heartening.  Understanding of how to replicate the results and select the intervention components or implementation models that are best suited to diverse settings are a priority for future impact on digital health equity.

### Acknowledgements

We acknowledge the important contributions of the clinicians, staff, and patients of the community health center who participated in the ACTIVATE project.

### Author Contributions

KK and DL acquired the funding for the study, conceived the project, and designed the study. KK oversaw technology development and intervention implementation. SM managed data curation and analysis. KK, DL, and SM contributed substantially to writing and they all edited and approved the manuscript for publication.

### Competing Interests

This study was funded by a gift grant from an anonymous, private donor who had no role in the study design or analysis.  SM and DL received salary support from the grant. KK was an employee of MITRE Corporation at the time of the project and received salary support from the grant. After completion of the study, the technology was spun out to Health Tequity LLC with the lead author serving as the Founder and CEO.

### Protection of Human Subjects

This study was conducted as a quality improvement project to enhance access and care management for patients with diabetes and hypertension served by the community health center.